\documentclass[final,twocolumn,showpacs,preprintnumbers,amsmath,amssymb,superscriptaddress]{revtex4}
\usepackage[utf8]{inputenc}
\usepackage{epsfig}
\usepackage{epstopdf}
\usepackage{amsmath,amssymb,amsthm}
\usepackage{dcolumn}
\usepackage{bm}
\usepackage{graphicx}
\usepackage{subfig}
\usepackage{caption}
\usepackage[]{natbib}
\usepackage{hyperref}

\begin{document}

\title{Optical response of $C_{60}$ fullerene\\ from a Time Dependent Thomas Fermi approach} 
\author{D. I. Palade}\email{dragos.palade@inflpr.ro}
\affiliation{National Institute of Laser, Plasma and Radiation Physics,
PO Box MG 36, RO-077125 M\u{a}gurele, Bucharest, Romania}
\author{V.Baran}\email{virbaran@yahoo.com}
\affiliation{Faculty of Physics, Bucharest}


\begin{abstract} 
We study the collective electron dynamics in $C_{60}$ clusters within the Time Dependent Thomas Fermi method in the frame of jellium model. The results regarding the optical spectrum are in good agreement with the experimental data, our simulations being able to reproduce both resonances from $20 eV$ and $40 eV$. We compare also, the results with those from other theoretical approaches and investigate the implications of quantum effects including exchange-correlation corrections, or gradient corrections from a Weizsacker term. The nature of the second resonance is studied using transition densities and phase analysis and interpreted as being a collective surface plasmon . 
\end{abstract}

\maketitle

\section{Introduction}
\label{intro}

The discovery of $C_{60}$ fullerene in 1985 \cite{kroto1985c} confirmed Buckminster's Fuller mental construction \cite{richard1965laminar} of a structure with high symmetries and outstanding stability properties. Since that moment a lot of attention has been concentrated on the special properties of fullerene, reflected in optical, electronic, thermal and mechanical \cite{dresselhaus1996science} behavior. The interest is not purely academical since the possibilities of using such clusters in biological \cite{bosi2003fullerene}, chemical and even cancer therapy \cite{mroz2007functionalized},\cite{mroz2007photodynamic} applications are actively discussed. For all this to become realistic, we need a solid understanding of the dynamical (optical) response of $C_{60}$ to external excitations as laser, projectiles, radiation, etc. Soon after the discovery of $C_{60}$, other fullerenes with lower symmetries have been observed and investigated, $C_{20}$, $C_{70}$, $C_{180}$, $C_{240}$, etc., but still, $C_{60}$ cluster remains the main figure in the this class.

One of the most intriguing and important (from an applicative point of view) properties of $C_{60}$ is related to its optical spectrum. Being a carbon compound, one could hardly view it as a metallic cluster or think about metallic properties. Even so, since the first theoretical calculations \cite{mroz2007photodynamic} and spectroscopic experiments \cite{hertel1992giant}, a giant resonance has been calculated and observed around $20 eV$ in its photoionization spectrum. It can be interpreted accordingly with the Mie's theory \cite{mie1908beitrage} as a surface plasmon which is usually present in metal nanostructures where the electrons have the characteristic of being quasi-free so the response function to external fields is caracterized by high values and a strong collective behavior. This aspect can be explained from the electronic structure $1s^22s^22p^2$ of carbon atoms, imagining that the valence electrons $2s^22p^2$ are quasi-delocalized in the cluster structure and capable of being collectively excited. Even peculiar was the presence of a second smaller peak in the optical spectrum around $40 eV$ revealed in photoionization experiments \cite{reinkoster2004photoionization},\cite{scully2005photoexcitation} both for neutral or ionized fullerenes. In the past decade, a lot of studies have reproduced it theoretically and interpreted it as a compressional collective mode associated with the so called \emph{volume plasmon}. Also, regarding its microscopical nature, pure quantum effects are thought to be responsible through a coherent superposition of orbital excitations. Some other studies, based on hydrodynamical pictures have argued that its nature is also that of a surface plasmon due to oscillations from the inner and outer surfaces of the electron cloud. Actually, one can find an impressive collection of papers regarding the existence and the nature of second resonance, but still there are debates about the correct interpretation.

While different methods, very involved from computational point of view, as Hartree-Fock \cite{sheka2007chemical},\cite{talapatra1992nonlinear}, RPA \cite{bertsch1991collective}, Density Functional Theory \cite{bauernschmitt1998experiment},\cite{scully2005photoexcitation},etc. have been used to study the optical spectrum in fullerene cluster, we intend in this paper to provide a study from the perspective of a semi-classical method, namely Time Dependent Thomas Fermi (TDTF). The method, or other variations of the same approximation, has been used before in connection with nuclei\cite{holzwarth1979fluid}, atoms \cite{horbatsch1981time}, clusters \cite{domps1998time}, strong laser interaction in clusters \cite{fennel2004ionization}, thin metal films \cite{crouseilles2008quantum} and even transport processes in semiconductors \cite{jungel2001quasi}. Therefore, it is eligible to consider such a method, which, to the best of our knowledge, has never been simulated before in its fully nonlinear form on $C_{60}$ fullerene.

Even though the approximations used are crude, the method has a set of advantages in respect with other more precise theories. First, as we shall see, it requires the propagation of a single pseudo-wave function in time in comparison with TDDFT or TD Hartree-Fock which in principle should propagate $240$ or even $360$ electron orbitals, and therefore, the gain in computational complexity and time is obvious. Second, the method allows us from its first principle to include or not, different corrections which account for pure quantum effects and so, we can investigate the level at which these effects become important in the dynamics. Third, the numerical scheme gives direct access to the local dynamics of electronic density, from which a quantitative description of the nature of present resonances is tractable. Besides all that, it is important to see, from numerical experiments, what are the capabilities of a semiclassical hydrodynamic theory in cases of such complex molecular systems as fullerenes are or what hopes are to use it in the future in other similar structures, at least to study the gross properties of dynamics.

\section{Theory}

\subsection{Jellium model and Thomas Fermi theory}
\label{TF and jellium}

In order to investigate the optical response of $C_{60}$ we do not need to involve strong excitations which could affect the core electrons or deform the structure and for this reason a simplification is at hand by using the so called jellium model. This model basically states that as long as in a molecular system the excitations are just perturbative and therefore lie in the linear region, one could say, due to Born-Oppenheimer \cite{born1927quantentheorie} approximation, that the nuclei remain frozen. In the same conditions, the core electrons are not excited and due to the electrostatic screening, the valence electrons, which have higher energies and are more susceptible to respond to external fields, see the coupling between nuclei and core in an averaged manner. So, instead of nuclei and bounded electrons it is more useful to use a single homogeneous distribution of positive charge which creates a fixed potential for the valence electrons. The self-consistent jellium model proved itself to be an appropriate method predicting quantitative results in good agreement with the experimental data in cluster physics \cite{brack1993physics}, \cite{de1993physics}. In our case, we use the classical model of a positive charged shell centered on the support surface for the carbon nuclei in $C_{60}$ cluster used before in other approaches \cite{yannouleas1994stabilized},\cite{rudel2002imaging}.

Going further to the basic approximation involved in the TDTF, we shall present first some features of the original, stationary, Thomas Fermi (TF) theory. 
In principle the approximation is derived considering an infinite system of non-interacting fermions. From the associated wave-functions, which are plane waves, the local density of kinetic energy is constructed and connected with the density of probability. This local form is used further in complex system with interactions to approximate also locally the true kinetic energy density. Besides the proper expression, it is useful to see how TF emerge as a genuine Density Functional Theory. It is known from Hohenberg-Kohn (H-K) theorems \cite{hohenberg1964inhomogeneous} that given a system of particles with a two-body interaction $U(\vec{r},\vec{r}')$ placed in an external potential $V(\vec{r})$, there is an unique functional of density called energy $E[\rho]$ which is minimized only by the true ground state density of the system $\rho_0(\vec{r})$:

\begin{equation}
E[\rho]=G[\rho(\vec{r})]+\int_{\mathbb{R}^3} \rho(\vec{r})[V(\vec{r})+\int_{\mathbb{R}^3}U(\vec{r},\vec{r}')\rho(\vec{r}')dr'^3]dr^3
\end{equation}

Where $G[\rho]$ is an universal (unknown) functional of density. In principle it should contain the kinetic energy of the system and other pure quantum effects as exchange-correlation. The variational principle of energy minimization must be applied with the constraint of constant number of particles $N$ so we call for the Lagrange multiplier technique:

\begin{equation}
\frac{\delta E[\rho]}{\delta \rho}=\mu
\end{equation} 

Where $\mu$ is the Lagrange multiplier interpreted as chemical potential. The associated Euler-Lagrange equation becomes (with the notation $\delta G/\delta \rho=g[\rho]$):

\begin{equation}
g[\rho(\vec{r})]+V(\vec{r})+\int_{\mathbb{R}^3}U(\vec{r},\vec{r}')\rho(\vec{r}')dr'^3=\mu
\label{EulerTF}
\end{equation}

Up to this level, the equation \ref{EulerTF} has been derived from the H-K only and the result is exact in the frame of DFT. Now, the mathematical form of TF theory states that $g[\rho]$ has the local form: $g[\rho]=\gamma\rho^{2/3}$ , with $\gamma=(3\pi^2)^{2/3}\hbar^2/2m$. 

Being derived under the homogeneous electron gas approximation, the functional, which in essence is an equation of state for the electron gas, works only in the case of systems with delocalized electrons and slow varying density (low gradients). Furthermore, it was proven that in molecular systems the TF theory is invalid since the atoms do not bind \cite{jahn1937stability} and the absence of exchange-correlation effects can give unphysical effects as can be the asymptotic behavior of the density. 

The drawbacks of the theory can be ameliorated by adding gradient corrections\cite{yang1986gradient} and exchange-correlation\cite{hodges1973quantum} terms in $g[\rho]$. Nonetheless, the method has been used before with good qualitative results in metal clusters \cite{kresin1988electronic},\cite{serra1989static} and even in $C_{60}$ \cite{palade2014general} and therefore our attempt can be justified, but we test both the original theory approximation and also the quantum and gradient corrected versions in order to make a comparison on the involved effects. 

\subsection{Derivation of Time Dependent Thomas Fermi equation}

As we shall see bellow, one of the form of TDTF consists of a hydrodynamical-like system of equations and it is natural to consider a derivation from the quantum Wigner equation within its Vlasov semi-classical approximation \cite{manfredi2005model}. Nonetheless, we shall derive it from Time Dependent DFT principles using the Lagrangian formalism just to have a better picture of the approximations involved and to be aware of the connections with TF theory. 

A similar short description can be found in \cite{domps1998time} designed for metal clusters. The main idea is to start, analogous with the stationary version (TF), from the quantum mechanical action and the variational principle of least action:

\[\mathcal{A}(t)=\int\limits_{0}^{t}\langle\Psi(\tau)|i\hbar\partial_{\tau}-\hat{H}|\Psi(\tau)\rangle d\tau
\]

Where due to a similar mapping with the one from K-S theorems, but proven by Runge and Gross \cite{runge1984density}, the state $\Psi$ is an unique functional of density: $|\Psi\rangle\equiv|\Psi[\rho(\vec{r},t)]\rangle$. $\hat{H}$ is the Hamiltonian operator. A set of hydrodinamical equations can be derived from the variational principle of least action \cite{runge1984density}:

  \begin{eqnarray}
  \label{cont}
  \partial_t\rho(\vec{r},t)+\nabla\vec{j}(\vec{r},t)=0\\
  \label{euler}
  \partial_t\vec{j}(\vec{r},t)=\vec{P}[\rho](\vec{r},t)\\
  \vec{P}[\rho](\vec{r},t)=-i\langle\psi[\rho]|[\hat{j},\hat{H}]|\psi[\rho]\rangle\label{dftpressure}
  \end{eqnarray}
  
In eq. \ref{dftpressure} $\hat{j}$ is the current operator while $\vec{P}$ is a three component vector. Unfortunately, the hamiltonian operator is not known explicitly and we can make approximations just on its expectation values, so there is no easy way to compute the commutator.

We start, as in \cite{lundqvist1983theory} with a fluid description of the system of particles analogous with a classical charged fluid with internal energy given by the energy functional from DFT. The dynamical part of the kinetic energy has the classical form described by a velocity field $\vec{u}(\vec{r},t)$. The expectation value of the Hamiltonian operator $\langle\psi|\hat{H}|\psi\rangle=\mathcal{H}[\rho,\vec{u}]$ is:

\begin{eqnarray}\nonumber
\label{semiclass}
\mathcal{H}[\rho,\vec{u}]&=&\int_{\mathbb{R}^3}\rho(\vec{r},t)[\frac{\vec{u}(\vec{r},t)^2}{2m}+\frac{\delta G[\rho(\vec{r},t)]}{\delta \rho}+\\&&+V(\vec{r},t)+\int_{\mathbb{R}^3}U(\vec{r},\vec{r}')\rho(\vec{r}',t)dr'^3]\mathrm{d}r^3
\end{eqnarray}

In order to compute the equations of motions from this approximation, we must introduce a scalar $S(\vec{r},t)$ as the conjugated variable of the density: $\vec{u}(\vec{r},t)\equiv\vec{u}[S(\vec{r},t)]$. Now we perform a Legendre Transformation $\mathcal{L}=\int_{\mathbb{R}^3}\rho\partial_tS-\mathcal{H}[\rho,S]$ from which the Lagrangian has been obtained and the equations of motion can be derived:

\begin{eqnarray}
\label{motion1}
\partial_t\rho(\vec{r},t)=-\frac{\delta \mathcal{H}}{\delta S}\\
\label{motion2}
\partial_tS(\vec{r},t)=\frac{\delta \mathcal{H}}{\delta\rho}
\end{eqnarray} 

Next approximation is the relationship between velocity field and $S$ field which can be adopted as: $\vec{u}=\nabla S$. Basically, the velocity field is taken to be irrotational , assumption which is not unrealistic, since our interest is the optical response of $C_{60}$ which is done in the usual way by studying the density response to a dipolar excitation, motion expected to have dipolar character and so, irrotational . With this representation and the notation $\frac{\delta G[\rho]}{\delta \rho}=g[\rho]$, the equations of motion \ref{motion1} and \ref{motion2} (which are at some level equivalent with \ref{cont}, \ref{euler}, \ref{dftpressure}) becomes:

\begin{eqnarray}
\label{s1}
   \partial_t\rho+\nabla(\rho\nabla S)=0\\
\label{s2}
   \partial_tS+\frac{|\nabla S|^2}{2m}+g[\rho]+V+U(\vec{r})=0    
\end{eqnarray}
 
The above set of equations can be embedded in a single non-linear Schrodinger like equation, using the so called Madelung transform \cite{madelung1927quantentheorie} defining the complex field $\Phi(\vec{r},t)=\rho(\vec{r},t)^{1/2}\exp(iS(\vec{r},t)/\hbar)$:  

 \begin{equation}
 i\hbar\partial_t\Phi(\vec{r},t)=-\frac{\hbar^2}{2m}\nabla^2\Phi(\vec{r},t)+w(\vec{r},t)\Phi(\vec{r},t)
 \label{eq:master}
 \end{equation} 

Where the pseudo-potential $w$ is described by:
  
 \begin{equation}
 w(\vec{r},t)=g[|\Phi(\vec{r},t)|^2]+V(\vec{r},t)+U(\vec{r},t)+\frac{\hbar^2}{2m}\frac{\nabla^2|\Phi(\vec{r},t)|}{|\Phi(\vec{r},t)|}
 \label{eq:potent}
 \end{equation} 
   
 Finally, setting $g[\rho]=\gamma\rho^{2/3}$, we obtain the TDTF model for the macroscopic dynamics of a system of identical particles in an external potential. In relation with the single pseudo-particle representation of Time Dependent DFT  $(\phi_i)$, namely the Kohn-Sham equations, those two approximations involved above, replace the true kinetic energy functional:
 \[T[\{\phi_i\}]=-\frac{\hbar^2}{2m}\sum\limits_{i=1}^{N}\int\phi_i^*\nabla^2\phi_idr^3\]
 
 with a picture in which, for the equation of state, all Kohn-Sham orbitals have constant amplitude $|\phi_i|=const$ and isotropic distributed $\nabla S^0_i$ $\forall$ $i=1,N$, while, for the dynamic part, all phases of $\phi_i$ have the same value $\arg(\phi_i)-\arg(\phi_i^0)=S$, $\forall$ $i=1,N$. Also, all the correlation and exchange effects are neglected.
 
 In the final step of interest, one should replace the mean field picture of interaction in the system with the Coulombian case: $U(\vec{r},\vec{r}')=|\vec{r}-\vec{r}'|^{-1}$.
 
 The result of this method is obvious: the final results (\ref{s1},\ref{s2}) are in fact a single NLSE, which even though works with a complex field which contains all the information that the hydrodynamical equations have, simplifies the numerical treatment with the capability of being tackled with specific and very efficient numerical methods as Cranck-Nicholson \cite{taha1984analytical} is. For this, it is known to be unconditionally stable and norm conserving, task which in the numerical treatment of hydrodynamic like system is achieved with greater computational costs, therefore, through this form we deal with the same problem but in a smarter way.

\subsection{Ground state and corrections}
\label{grndst+correct}

It is intuitive to think that the stationary solution of eq. \ref{eq:master} is the same with the TF eq. \ref{EulerTF}. This indeed can be proven with the \emph{ansatz} $\Phi(t,\vec{r})=\Phi_0(\vec{r})e^{i\mu t/\hbar}$ which gives us from the equation \ref{eq:master} an eigenvalue problem :
   
   \begin{equation}
   \mu\Phi_0(\vec{r})=[-\frac{\hbar^2}{2m}\nabla^2+w(\Phi_0(\vec{r}))]\Phi_0(\vec{r})
   \label{eq:eigenvalue}
   \end{equation}
   
   Or an equivalent Euler-Lagrange (TF) equation:
   
\begin{equation}
\mu=g[|\Phi_0(\vec{r})|^2]+V(\vec{r})+\int\frac{|\Phi_0(\vec{r})|^2}{|\vec{r}-\vec{r'}|}d\vec{r'}
\label{eq:tf}
\end{equation}

The numerical methods of solving this problem are known and will be discussed in detail later.
   
Now that the basic approximations of TF and TDTF are clear: a) an equation of state for the electronic gas approximated locally with the one from the homogeneous infinite model, b) a decoupled kinetic energy in the state energy and a irrotational flow kinetic energy, it is important to analyze what do these approximations miss and how could they be improved.

First of all, starting from the Wigner-Kirkwood expansion \cite{fujiwara1982wigner} of Bloch density function, one derives TF equation of state just as an $0^{th}$ order approximation which works only for very slow varying densities, therefore, supplementary higher order (gradient) corrections could be added. This is done trough a Weiszacker term $\lambda (\nabla^2\rho^{1/2})/\rho^{1/2}$ with $\lambda=1/8$ \cite{weizsacker1935theorie} which should account, in principle, for local oscillations of the density profile \cite{yang1986gradient}.

Second, the universal functional $G$ does not account only for the kinetic energy of the system but for the exchange-correlations also. For that reason we can correct $g[\rho]$ adding a supplementary term present in the DFT potential of the Kohn-Sham equations. With all this corrections the new local density approximation can be written :

\begin{equation}
g[\rho]=\gamma\rho^{2/3}+\lambda\frac{\hbar^2}{2m}\frac{\nabla^2\rho^{1/2}}{\rho^{1/2}}+v_{xc}[\rho]
\end{equation}
\begin{eqnarray}\nonumber
\label{gun}
v_{xc}(\rho)&=&-\frac{e^2}{4\pi\varepsilon_0}(\frac{3}{\pi})^{1/3}[\rho^{1/3}+\\&&+\frac{0.054}{a_0}ln(1+11.4a_0(\frac{4\pi\rho}{3})^{1/3})]
\end{eqnarray}

The $v_{xc}$ potential can have in principle any form used in general DFT calculations, but in practice we choose the Gunnarson-Lundqvist functional from equation \ref{eq:gun} which has been used before succesfully in cluster dynamic simulations \cite{feret1996electron}, even if it was designed, in principle, for ground state calculations.
 
Some applications \cite{domps2000semi} in metal clusters with the semi-classical Vlasov method have used only the $v_{xc}$ correction. But following \cite{bonitz2004quantum} we will include both Wiezsacker and xc, or none, since from kinetic perspective, both gradient corrections and exchange correlation effects have the same magnitude and therefore, spurious effects could arise if they are not used in pair.

\section{Results and discussion}

\subsection{$C_{60}$ ground state}

Going back to $C_{60}$, we shall start by investigating the properties of electronic system in the ground state as describe by TF theory. This is done by solving eq. \ref{EulerTF}. The jellium model is used to describe the coupling between nuclei and core electrons ($1s^2$ shell) in a homogeneous positively charged shell centered on the spherical surface that contains the geometric position of nuclei $\rho_{jel}(\vec{r})\propto \Theta[(r-r_1)(r_2-r)]$. This sphere has a radius of $r_0=3.54 \mathring{A}$ and the thickness is taken as in \cite{rudel2002imaging}, \cite{puska1993photoabsorption},\cite{weaver1991electronic} $\Delta \approx 1.5 \mathring{A}$, therefore, $r_1\approx 2.8\mathring{A}$ and $r_2\approx 4.2\mathring{A}$.

\begin{figure}[h]
\includegraphics[width=0.48\textwidth]{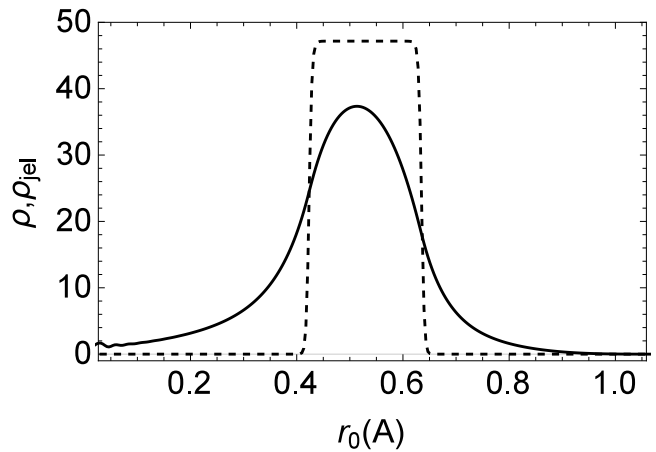}
\caption{Radial profile of jellium density (step-like, dashed) and radial profile of electron density (continuous) in $C_{60}$ obtained with TF method}
\label{fig:grdens}
\end{figure}

One of the most natural way of solving eq. \ref{EulerTF} is to start with an initial guess on ground state density $\rho_0$ from which the Coulomb potential is constructed with the help of Poisson equation $\nabla^2 \phi=-e\rho/\varepsilon_0$ and then from $g$ functional, the new density and the chemical potential are calculated in such a way that the normalization condition $\int\rho (\vec{r})dr^3=240$ is fulfilled. The process is iterated until the convergence it is reached.

Since in the case of gradient and exchange-correlation correction, solving the equation both for 
$\rho$ and $\mu$ can be problematic, we have chosen instead, to solve the eigenvalue problem \ref{eq:eigenvalue}. In a similar way, we start with a guessed $\Phi_0(\vec{r})$ from which the potential $w(\vec{r})$ is constructed and then, in a finite difference description on real space of the laplacian operator, the spectrum of eigenvalues and eigenvectors is computed. The eigenvector with the lowest eigenvalue is taken as new solution and the loop is repeated until convergence. In this way, the $g$ functional can be computed directly from the last solution and it is not neccesary to solve complicated equation as $g(x)=a$. 

\begin{figure}[h]
\includegraphics[width=0.48\textwidth]{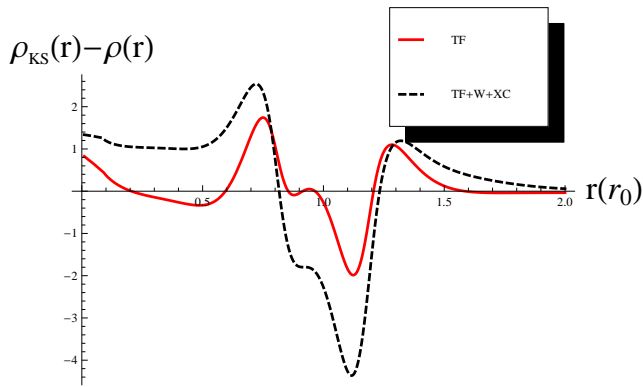}
\caption{Difference between the radial averaged profile of electron density obtained in DFT calculations and the one obtained with TF method and TFWXC corrected version}
\label{tfdftt}
\end{figure}

The results in the electron density are physical and in good agreement with the experimental values for the inner, $\simeq 1.8 \mathring{A}$ and outer radius of the fullerene $\simeq 5.1 \mathring{A}$, see Fig. \ref{fig:grdens}. For comparison we have performed also DFT calculations with the same jellium model, and solved the Kohn-Sham equations using the Gunarson-Luidqsvit potential (\ref{eq:potent}). Also, in Fig. \ref{tfdftt}, the difference between the radially averaged Kohn-Sham density and the Thomas Fermi one is plotted as a quantitative description of how close to DFT-reality is TF ground state in the case of fullerene.  The result is that local errors stay under $10\%$ which is an unexpected result regarding the simplicity of the model. Nonetheless, this is in part a consequence of the smooth of jellium.

\subsection{Numerical details}
\label{numeric}

From numerical point of view, we have chosen to solve both the stationary eq \ref{eq:master} and the time dependent one \ref{eq:eigenvalue} in real space under the finite difference approximation. The ground state is taken as being spherically symmetric with the only variable involved being the radial one, so very refined grids have been used. For example, a radial domain of $0<r<3r_0$ divided in $n=1000-10000$ equidistant points has been employed usually which gave a very accurate picture for both Coulomb potential and the more stiff terms as Bohm potential or the correlation potential. Convergence has been achieved usually within $200-300$ iterations with a criterion on the solution norm of $||\Phi_0^{{k+1}}-\Phi_0^{{k}}||<10^{-4}$.

Considering the spherical symmetry of the ground state and the azimuthal symmetry of the excitation mechanism (see bellow) all this coupled with Schrodinger aspect of the eq. \ref{eq:master}, it becomes obvious that $\Phi(\vec{r},t)$ is also angular symmetric. Therefore, we use further a cylindrical coordinate system with angular symmetry, keeping the 2D dependence only in $(r,z)$ coordinates, where $-3r_0<z,r<3r_0$. Indeed, this is  an overestimation of the box which contains the entire electron distribution, but is was necessary in order to prevent false fractional electrons escaping from the cluster. Also, it is important to use such large distances as boundaries of the box since the Coulomb potential is solved with Poisson equation which needs a set of boundary conditions, extracted from a multipole expansion, valid only in the large distance approximation. 
 
The radial coordinate has been taken to cover the whole diameter, as described in \cite{mohseni2000numerical} to avoid the singularity from $0$ and the implementation of boundary conditions for Poisson equation in $r\approx 0$ which had have required a supplementary, unnecessary effort. 

Usual discretizations involved a number of points of $nr=nz=100-200$ on each axis. We have found that this values offer a good precision-computational cost ratio and deals with the possible stiffness of the potential. 

Regarding the time propagation, we have used a total time of propagation around $1-50 fs$ enough for the collective motion to fade through dissipation and to avoid instabilities. This interval has been discretized with a step between $\delta t=10^{-3}-10^{-2}fs$. The numerical method invoked in the actual propagation of $\Phi$ was the Crank-Nicholson (CN) which assures us stability and norm conservation. Because we have self-consistency between the mean field potential $w$ and $\Phi$, we use a mid-point rule to avoid numerical energy losses. We use $w[\Phi]$ to propagate a false solution $\Phi^*$ then, we reconstruct a mid-point potential $(w[\Phi(t)]+w[\Phi^*])/2$ which will propagate $\Phi(t)$ to $\Phi(t+\delta t)$.

Regarding the excitation mechanism, the most natural would be to consider a weak laser pulse described classically (due to size of the cluster, an external electric field is fairly approximated to a plane wave) as $v(\vec{r},t)\propto cos(\omega t-kz)e^{-\eta t}$. Nonetheless, this choice can induce $\omega$ or $k$ dependent solutions which is undesirable. Therefore, we take as initial condition a dipole shift ground state profile, $\Phi(r,z,t=0)=\Phi_0(\sqrt{r^2+(z+\eta)^2})$ as in \cite{calvayrac1997spectral}. Usual choices on $\eta$ have been $\eta=1-10\%r_0$.

Solving eq. \ref{eq:master} in the numerical frame described above, the single viable result is the density distribution $\rho(\vec{r},t)=|\Phi(\vec{r},t)|^2$. We are interested in the optical response so we should investigate the cross-section dependence with energy. But since cross-section is closely related to the strength function, we compute in our simulations the Fourier transformed of the total dipolar moment:

\[\mathcal{D}(\omega)=\int e^{i\omega t}d(t)d t\]

\[d(t)=\int z\rho(\vec{r},t)dr^3\]

which is tackled with the Fast Fourier Transform algorithm.

\subsection{Validity and correction improvements}

As discussed in the derivation of TDTF, the approximations involved are the TF local form of the equation of state and a single valued irrotational velocity field $\vec{u}$ associated with the dynamics of electrons. In the picture of phase space described mathematically by the Wigner function $f_w(\vec{r},\vec{p},t)$, the model reduces to a geometrical picture in which $f_w$ has a constant value in a sphere displaced accordingly with $\vec{u}$ : $f(\vec{r},\vec{p},t)\propto\Theta[|\vec{p}-(\vec{p}_F+\rho\vec{u})]$. This approximation should underestimate from the start any collisional effects and the velocity dispersion in momentum space due to Fermi sphere anisotropies, so we expect that the dissipation in the system will be diminished.

\begin{figure}[h]
{\includegraphics[width=0.48\textwidth]{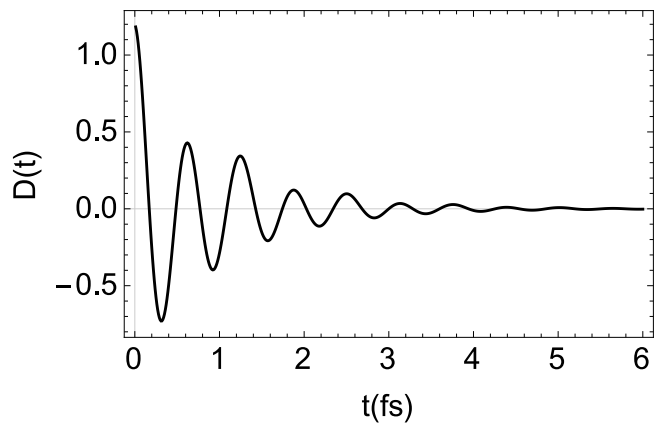}}
\caption{Total dipole moment through first $2fs$ of dynamics}
\label{Fig:2a}
\end{figure}

\begin{figure}[h]
{\includegraphics[width=0.48\textwidth]{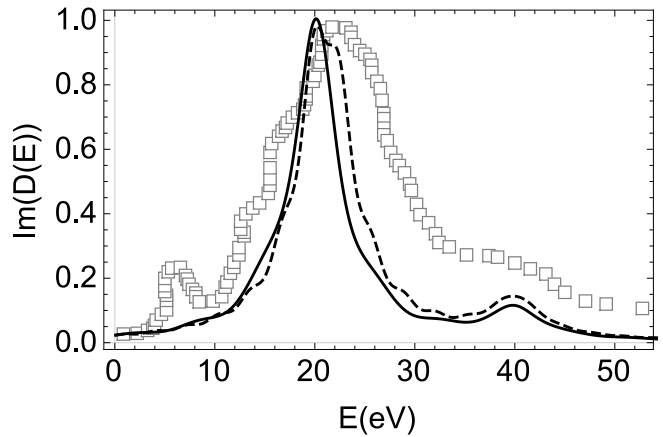}}
\caption{Experimental (squares)\cite{p2008absolute} and theoretical (TF-full line, TF-WXC- dashed) absorption spectrum}
\label{Fig:2c}
\end{figure}

The time evolution of the dipolar moment is plotted in Fig. \ref{Fig:2a} in which the damping of oscillations can be observed. Still, this is mathematically reproduced due to non-linearities of the effective potential \ref{eq:potent} and not from the above mentioned effects which are missed. A qualitative image on how much damping is not reproduced by this model is in Fig. \ref{Fig:2c} where the experimental spectrum and the one obtained in the present work are reproduced. Both, the width of the resonances and the sum rule are underestimated.

Further more, the basic model used for picturing nuclei and core electrons was that of a jellium model. This aspect is not neglijable and it is at some level a surprise that such a complex cluster with icosahedral geometry like fullerene does not present a rough effect on the position of resonances due to ionic structure. Other methods have obtained a shift in the position of plasmons and this defect has been corrected with a shift back of almost $5eV$ \cite{brack2008coupling} explained exactly as ionic effect. In our simulations, we have shifted the spectrum with $2eV$ to fit the experimental peak from $20eV$.

Many works, as \cite{scully2005photoexcitation} which investigate the dynamics with TDLDA interpret the second resonance as a collective phenomenon. We confirm this interpretation based on the fact that our method has no single particle effects therefore, any oscillation captured in the calculations can have only collective behavior.

Moreover, wanting to simulate our share of quantum effects, we have added to dynamics the supplementary exchange-correlation potential (LDA) from \cite{hedin1971explicit} coupled with the gradient correction \cite{yang1986gradient} and, as it can be seen from \ref{Fig:2c} no notable changes are obtain in the spectrum.

\subsection{Plasmons: existence and interpretation}

We argue that in usual spherical clusters, $Na$ for example the resonance at Mie frequency appears naturally due to their geometry and the oscillations at the surface. In $C_{60}$ we have two connected surfaces, one outside $r\approx 1.3r_0$ and one inside $r\approx 0.7 r_0$ the cluster. Each one of them has specific resonances, which reconstruct the total optical response. Does this exclude the interpretation of volume plasmon for the $40eV$ resonance?
To answer to this question we should follow the debate \cite{korol2007comment}-\cite{PhysRevLett.98.179602}. While \cite{korol2007comment} argues based on a linearized hydrodynamic model, that the resonances near $22eV$ and $38eV$ must both be associated with a surface plasmon excitation. On the other side, in the response of \cite{PhysRevLett.98.179602} the authors consider the problem only one of terminology, given the fact that indeed there are two resonances on the surface of the electron cloud in fullerene, but it is interpreted as a \emph{volume plasmon} due to the modulations of the electron density inside the system.

We want to investigate the spatial position of the resonances and, if possible, to see how does the inner density modulates with time in order to confirm one of the physical pictures of the above. For this we will work with transition densities in our system. While in a stationary problem, as the static polarizability is, the representation of transition density $\rho(\vec{r})-\rho_0(\vec{r})$ is straightforward to do, in a time-dependent system, becomes hard (and useless) to see the time dependency of the same quantity. Therefore, we look only on the spectral amplitude of response defined as the modulus of:
	 
	 \[\chi(\vec{r},\omega)=\int\limits_{0}^{\infty}\rho(\vec{r},t)e^{i\omega t}dt\]
 
\begin{figure}
\includegraphics[width=0.48\textwidth]{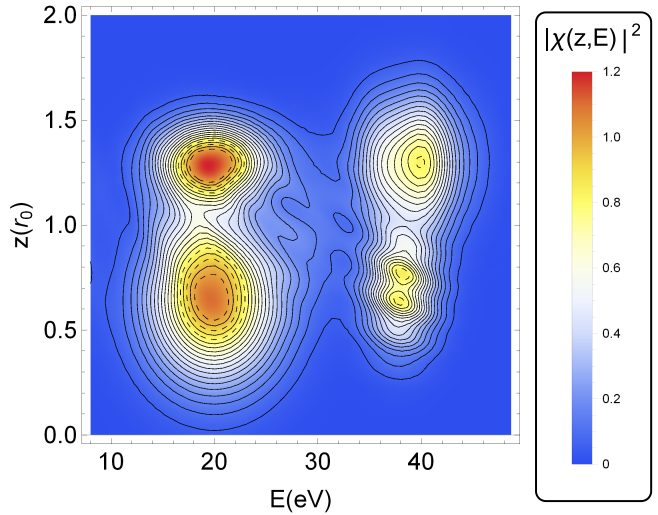}
\caption{Transition densities in the $\hbar\omega$-$z(r_0)$ space}
\label{space}
\end{figure}

Also, we chose for a visual representation only an azimuthal profile ($z=0$) to look at $|\chi(r,z=0,\omega)|^2$. Further, we set for this a reasonable description of the \emph{volume} domain in the structure, as the spherical shell defined as $70\%r_0<r<130\%r_0$, where, from numerical simulation of the ground state, around $90\%$ of the total electronic charge it is contained. Therefore, we can interpret that the adjacent space of this shell can be considered \emph{surface} of the system. In Fig. \ref{space} the results can be seen in a density plot. As one expected, high peaks of \emph{activity} are present around $20 eV$ and smaller one around $40eV$, located spatially on what we have defined as surfaces of the system. What is important to look at is the activity in the inner region, the volume one, where there is a non-zero distribution of amplitude. 

If there have been no activity we could have argued that during the oscillation we could have a divergent-free current field in this region which would allow accordingly with the continuity equation $\partial_t \rho = - \nabla \vec{j}$ to have almost null oscillation in density. But as the amplitudes are positive in the volume of fullerene, we support the idea of modulated density and the fact that correct physical picture is indeed that of a volume plasmon as two surfaces resonances which modulate the volume electron density.

Further, the oscillations are not in phase and this supports once more the above interpretation.

\section*{Conclusions}

Starting from the fundamental theorems of DFT and TDDFT we have presented a derivation of the so called TDTF method which uses as approximations the TF equation of state and the picture of single valued velocity field for a fermionic system. The semi-classical and exchange-correlation corrections are discussed and included for comparison in our simulations. The method is used in the case of dipolar oscillations in $C_{60}$ cluster. Under spherical symmetry and jellium model approximation, the electronic ground state radial profile is calculated in good agreement with the experimental descriptions of the inner and outer surfaces. 

Furthermore, in the dynamical regime, we excite the electrons by a small initial dipole shift and let the system to evolve under the TDTF equation in its NLSE form. We found a good result in the optical spectrum where the resonance from $20eV$ and that from $40eV$ are semi-quantitatively obtained. The inclusion of corrections is found to produce no essential differences from the pure TF approximation. 

In order to have a quantitative description of the second resonances we analyze the local transition densities from which we conclude that in $C60$ optical response we have two out-of-phase surface oscillations which account both for the $20eV$ peak and also for the $40eV$ one, due to the induced compression of the electron density in the volume of the system. The problem is indeed one of terminology, while the physical picture is clear and confirmed by our analysis.

We intend in future studies to test the capabilities of this method in other complex systems, probably the new discovered boron bucky-balls\cite{zhai2014observation} which have a similar structure, but more refined jellium model should be used.

\bibliographystyle{unsrt}
\bibliography{bibc60tdtf}

\end{document}